\documentclass[twocolumn,prl,showpacs]{revtex4-1}
\usepackage{graphicx,color}
\usepackage{dcolumn}
\usepackage{amsmath,amssymb}
\usepackage{bm}

\newcommand{\Dy}{Dy$_2$Ti$_2$O$_7$}
\newcommand{\Ho}{Ho$_2$Ti$_2$O$_7$}
\newcommand{\kB}{k_{\rm B}}
\newcommand{\etal}{{\it et al.}}
\newcommand{\LpMax}{67,108,864} 
\newcommand{\LowestT}{0.33} 

\newcommand\zuPYROSG{%
  \begin{figure}[t]
   \includegraphics[width=0.41\linewidth]{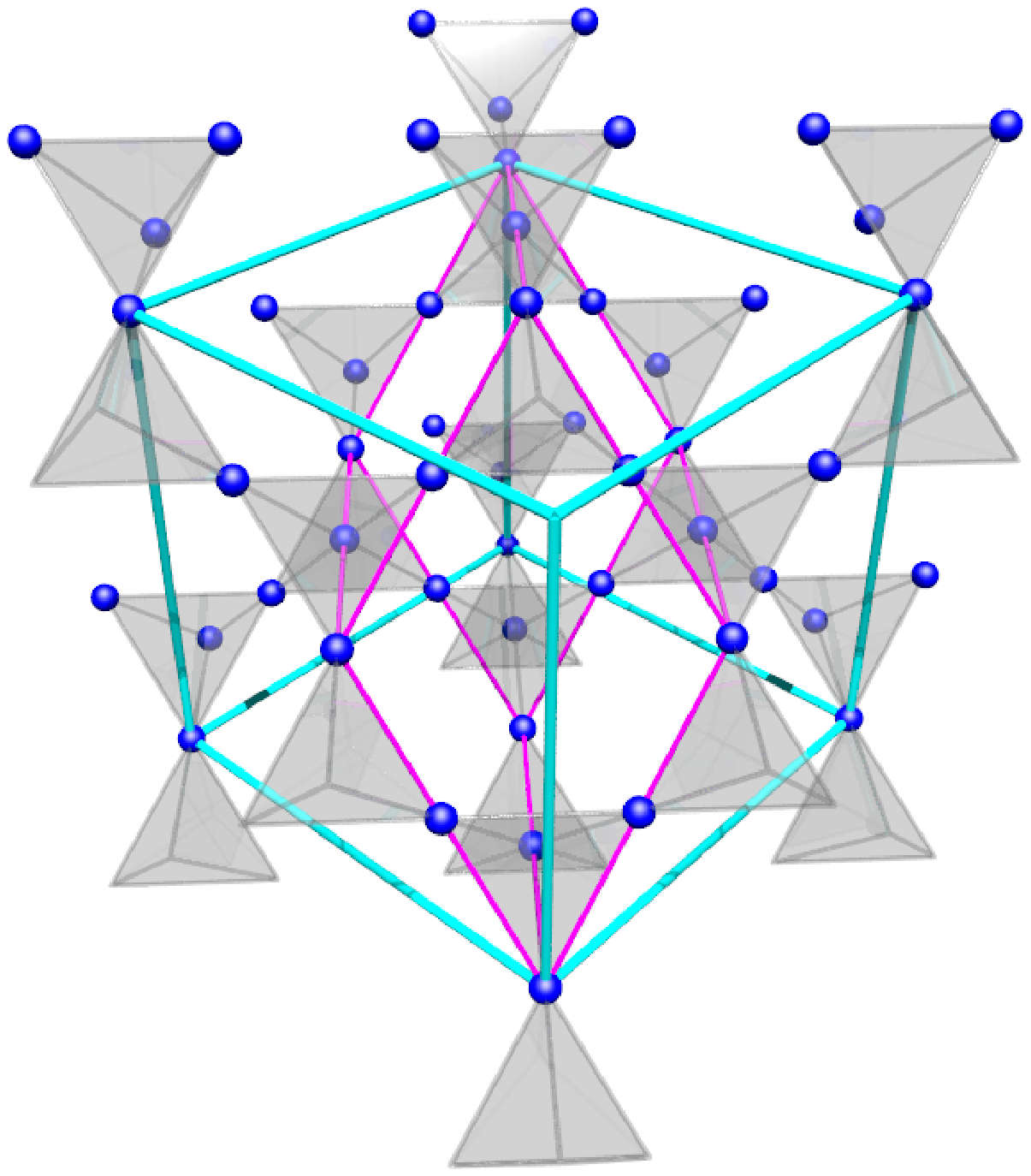}\hfill
   \includegraphics[width=0.56\linewidth]{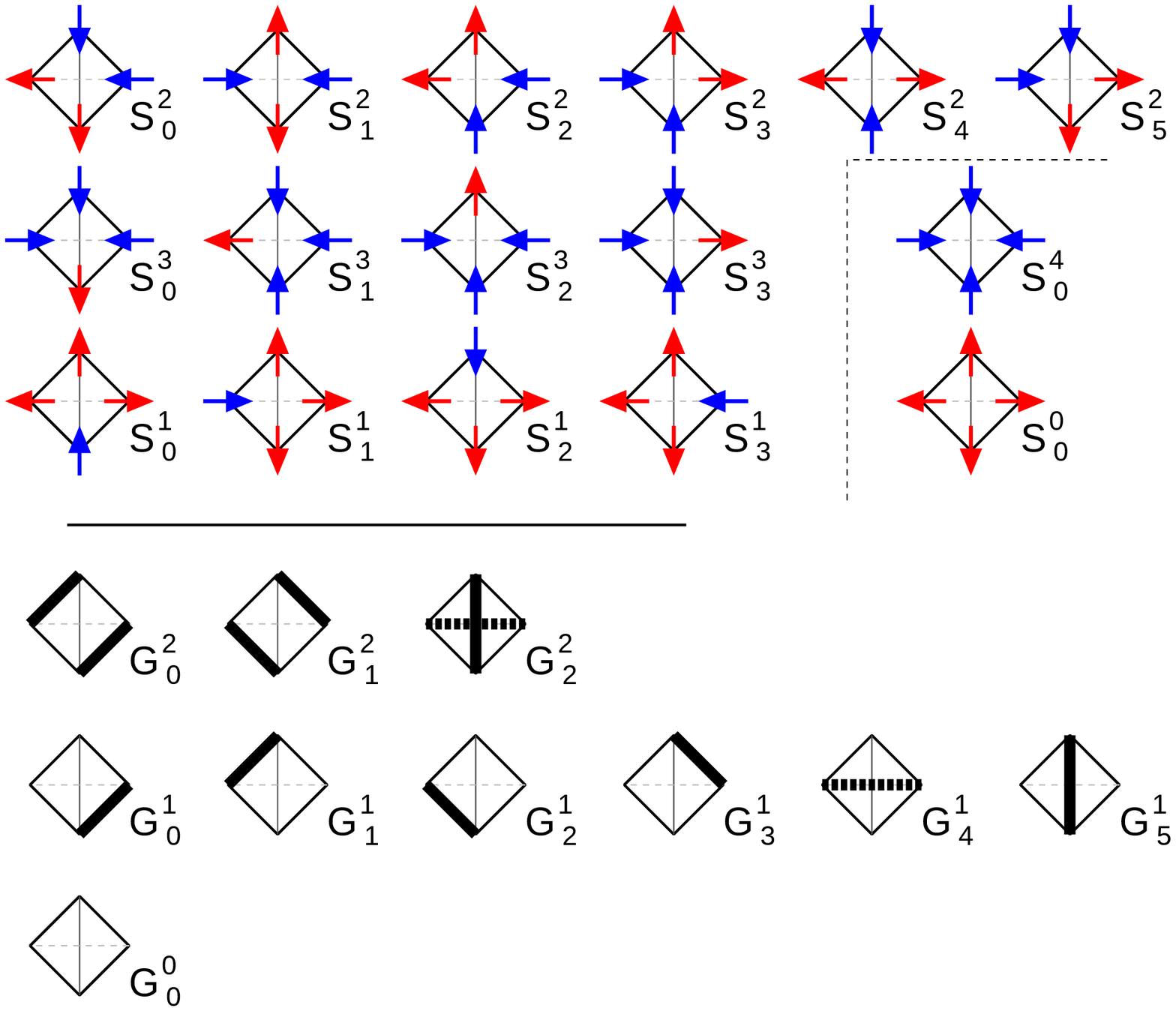}
   \caption{%
   (Color online)
   Left:  
   The pyrochlore lattice $\Lambda_{\rm p}$. 
   Blue dots represent magnetic ions.
   The frame in magenta (cyan) marks the primitive (cubic) cell with 4
   (16) spins. 
   The edge length and fundamental vectors of the cubic cell are,
   respectively, $a$ and $({\bf e}_x,{\bf e}_y,{\bf e}_z)$.
   The length of primitive vectors, e.g.,
   ${\bf t}_1=\frac{a}{2}({\bf e}_x+{\bf e}_y)$
   is $a/\sqrt2$.
   Right: 
   Top three lines give the 16 states per one tetrahedron $S_u^\mu$
   classified by the number of inward spins $\mu$ (blue arrows).
   Bottom three lines give the 10 graphs $G_v^\nu$, where $\nu$
   denotes the number of bonds (bold lines).
   }
   \label{PYROSG}
  \end{figure}
}

\newcommand{\figurescalelarge}{0.80}

\newcommand\zuSDT{%
  \begin{figure}[t]
   \includegraphics[width=\figurescalelarge\linewidth]{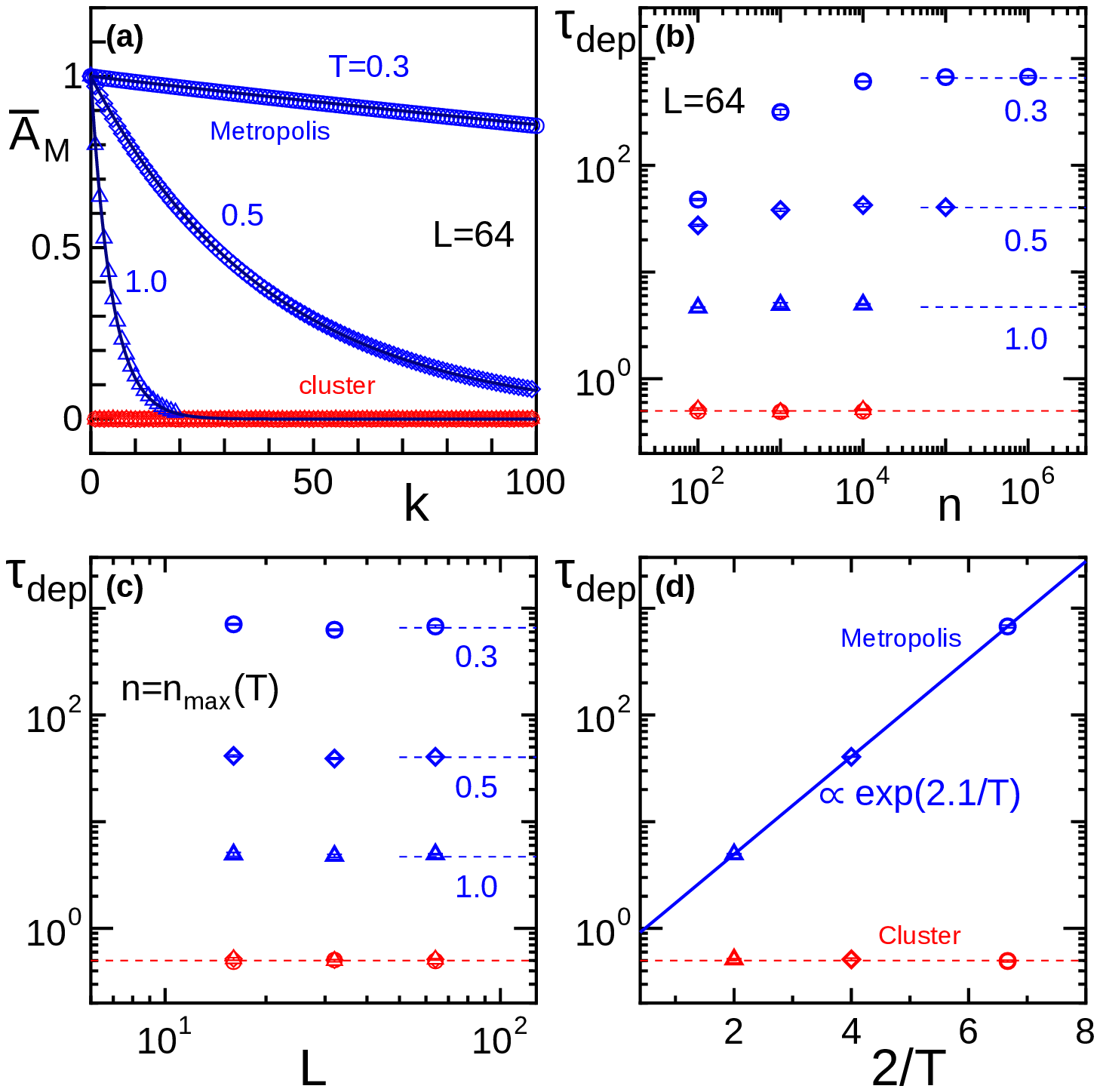}
   \caption{%
   (Color online)
   (a): 
   $\bar A_{M}(k)$ at $T=0.3$, 0.5, and 1.0. 
   Blue (red) marks plot data by the Metropolis (cluster) algorithm. 
   The forms of exponential relaxation are drawn as fitted curves in
   blue.
   (b) and (c):
   The $n$ and $L$ dependencies of $\tau_{\rm dep}$ in two algorithms.
   The relaxation times estimated via $\bar A_{M}(k)$ are given as
   blue dotted lines; the red indicate $\tau_{\rm dep}=\frac12$. 
   (d): 
   Temperature dependence of the relaxation times; 
   the fitted line shows its exponential dependence in the Metropolis
   dynamics. 
   }
   \label{SDT}
  \end{figure}
}

\newcommand\zuPP{%
  \begin{figure}[t]
   \includegraphics[width=\figurescalelarge\linewidth]{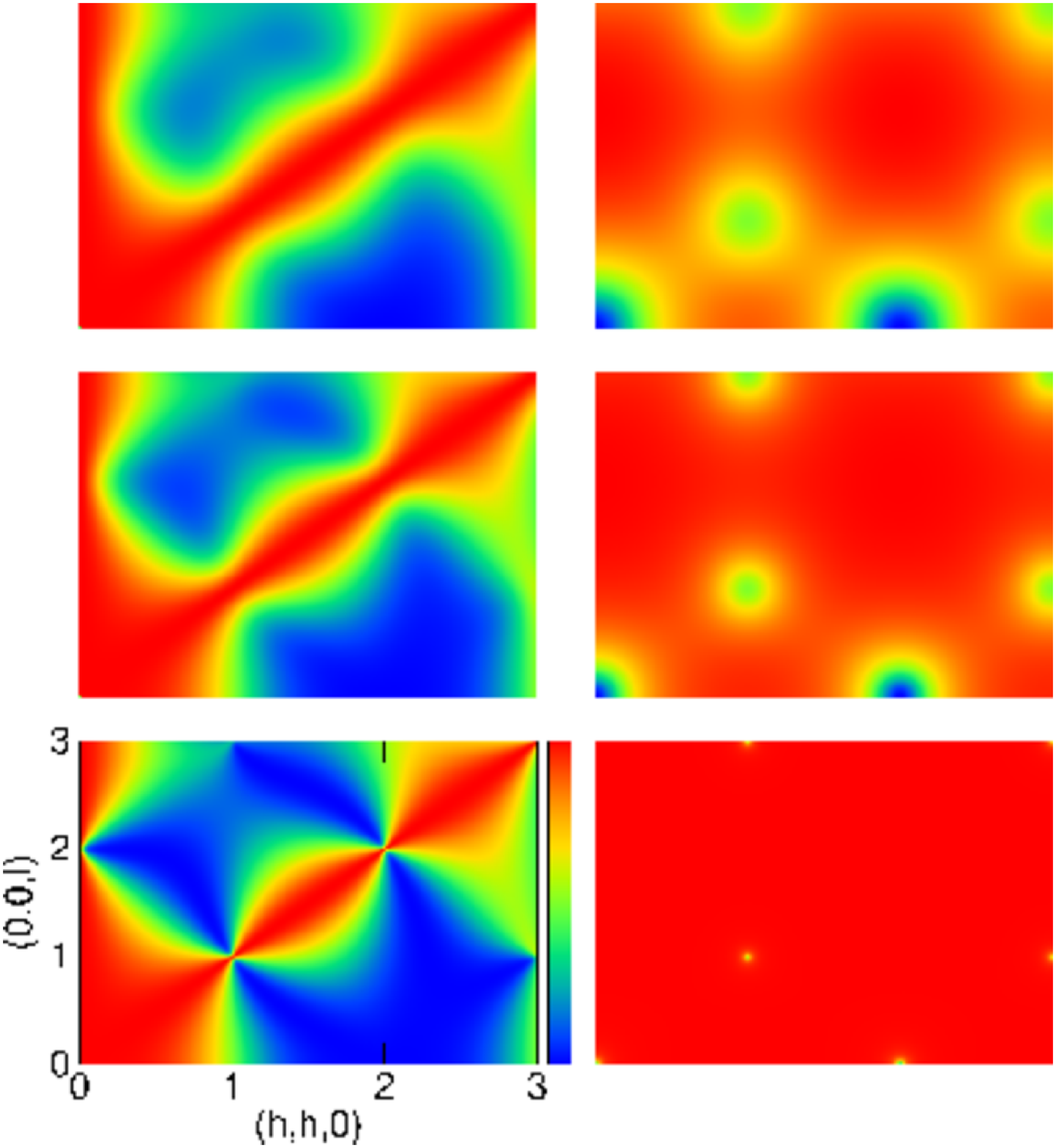}
   \caption{%
   (Color online)
   Color maps of spin and charge structure factors,
   ${\cal S}_{\perp}({\bf Q})$ and ${\cal C}({\bf Q})$,
   at
   $T=$0.3 (the bottom), 1.0 (the middle), and 2.0 (the top).
   The data in the left (right) panels are normalized by an amplitude of
   spin (charge) structure factor at each $T$.
   The wavevector is on the $(h,h,l)$ plane, and is measured in units of
   $2\pi/a$.
   }
   \label{PP}
  \end{figure}
}

\newcommand\zuFSING{%
  \begin{figure}[t]
   \includegraphics[width=\figurescalelarge\linewidth]{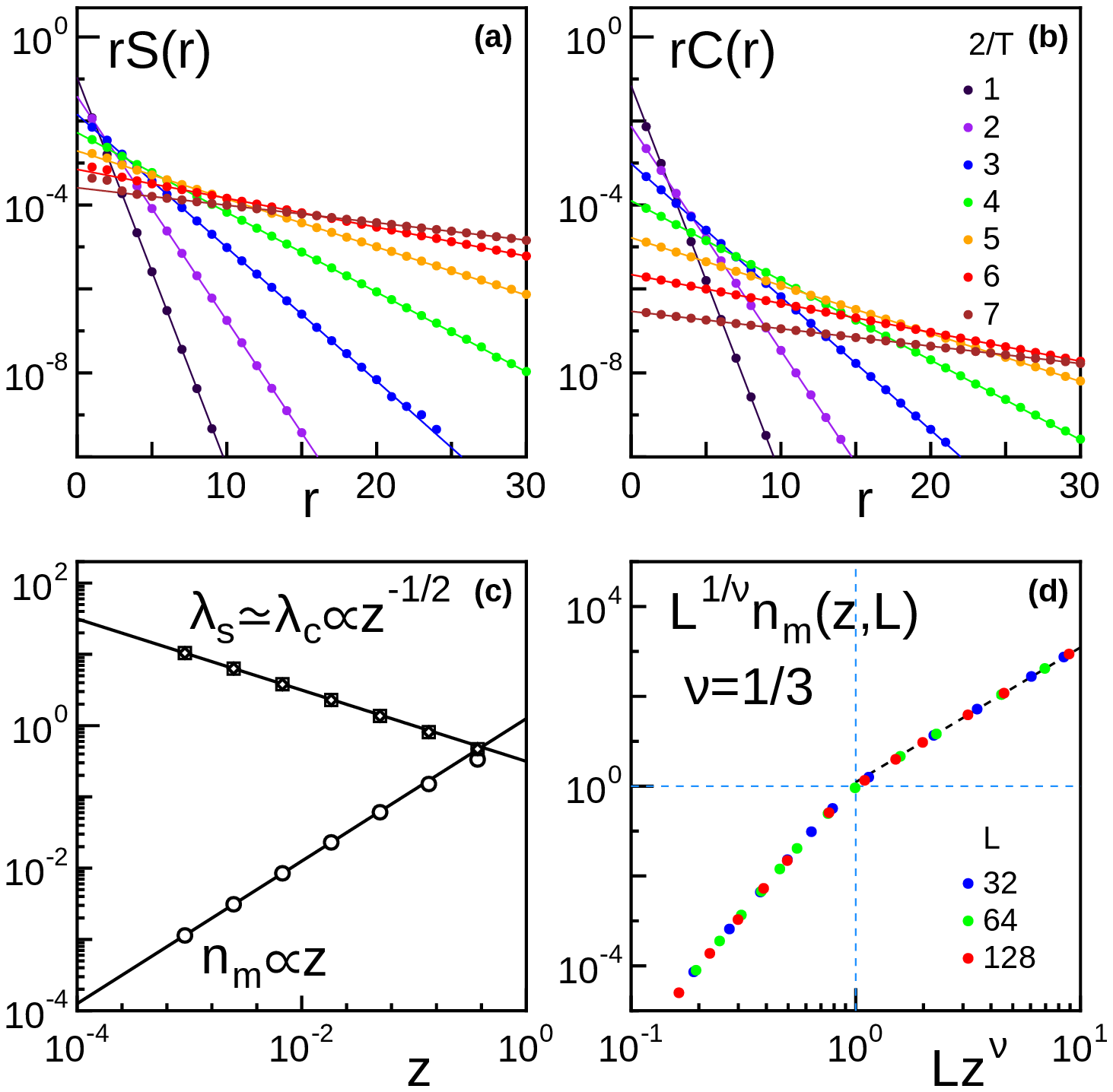}
   \vspace{-2mm}
   \caption{%
   (Color online)
   (a) and (b): 
   $r{\cal S}$ and $r{\cal C}$ versus $r$ for various $T$.
   Slopes of fitted lines determine the inverse correlation lengths
   for the $L=128$ system.
   (c):
   The $z$ dependence of
   $\lambda_{\rm s}$ (diamonds),
   $\lambda_{\rm c}$ (squares), and
   $n_{\rm m}$ (circles).
   The slope of the line for $\lambda_{\rm s,c}$ is $-\frac12$,
   and for $n_{\rm m}$ is 1.
   (d)
   gives the finite-size-scaling plot of $n_{\rm m}$ with the critical
   exponent $\nu=\frac13$.
   \label{FSING}
   }
  \end{figure}
}

\begin{document}
\title{A cluster algorithm for Monte Carlo simulations of spin ice}
\author{Hiromi Otsuka}
\affiliation{Department of Physics, Tokyo Metropolitan University, Tokyo 192-0397, Japan}
\date{\today}
\begin{abstract}
 We present an algorithm for Monte Carlo simulations of a
 nearest-neighbor spin ice model based on its cluster representation.
 To assess its performance,
 we estimate a relaxation time, and find that, in contrast to the
 Metropolis algorithm, our algorithm does not develop spin-freezing.
 Also, 
 to demonstrate the efficiency, we calculate the spin and charge
 structure factors, and observe pinch points in a high-resolution color
 map.
 We then find that
 Debye screening works among defects and brings about short-range
 correlations,
 and that
 the deconfinement transition triggered by a fugacity of defects $z$ is
 dictated by a singular part of the free-energy density
 $f_{\rm s}\propto z$.
\end{abstract}

\maketitle

 The pyrochlore lattice
 $\Lambda_{\rm p}$
 (Fig.\ \ref{PYROSG}) is one of the more
 important structures in condensed matters physics
 \cite{Bramwell01}
 because inherent frustrations can inhibit the stabilization of
 conventional phases, and also not a few numbers of pyrochlores have
 been synthesized and have provided measurements of their static and
 dynamical properties.
 The so-called spin ice materials {\Ho}
 \cite{Harris97}
 and {\Dy}
 \cite{Ramirez99}
 are examples; 
 despite their large magnetic moments, they were found not to be
 magnetically ordered, but to exhibit a residual entropy.
 Moreover,
 with the intriguing proposal of magnetic monopolelike excitations in
 these materials
 \cite{Castelnovo08}, 
 intensive efforts have been paid to account for their unconventional
 phases and exotic excitations near the ground states
 \cite{Hertog00,Bramwell01PRL,Melko01,Ryzhkin05,Jaubert08,Fennell09,Fennell09,Castelnovo11,Jaubert11,Jaubert13,Sen13,Otsuka14}.

 In general,
 numerical approaches can play an important role in this developing
 research area;
 indeed, efficient methods have also been sought for spin ice.
 The Monte Carlo (MC) method is apparently of principal importance,
 but the single-spin flip dynamics based on the Metropolis algorithm is
 inefficient at lower temperatures due to spin-freezing
 \cite{Melko01}.
 This indicates that
 we should employ spin-flip dynamics compatible with the ice rule
 \cite{Bernal33},
 which we shall soon discuss.
 In studies of vertex models for water ice,
 Barkema and Newman closely discussed multi-spin-flip algorithms using
 loop clusters of spins
 \cite{Barkema98}.
 Based on their discussion,
 Melko {\etal} introduced loop flips into the Metropolis algorithm, and
 succeeded in simulating a spin-ice model at lower temperatures
 \cite{Melko01}.
 Jaubert {\etal} re-expressed the local Boltzmann weight as a sum of
 graph weights, and constructed a loop cluster which can be freely
 flipped (worm algorithm)
 \cite{Jaubert08}.
 Also,
 more resent research reported loop-update MC studies on other types of
 pyrochlore magnets
 \cite{Shinaoka2010}.

 In this rapid communication
 we propose a MC algorithm for a simple spin-ice model based on its
 cluster representation.
 Significantly,
 the algorithm is quite efficient in simulations for all temperatures,
 and thus provides us an opportunity to simulate large systems very
 accurately; indeed this is necessary for studies on phase transitions
 and critical phenomena as we see below.
 Also,
 our algorithm is simple in that it could provide a building block for
 simulations of more complicated systems including dipolar spin ices.
 In what follows,
 we shall detail our algorithm and provide numerical results.

 \zuPYROSG

 {\it Model:}
 Suppose that
 $p$ specifies a site in
 $\Lambda_{\rm p}$
 and that
 the unit vectors for the local axes $\hat{\bf u}(p)$ and the Ising
 variables $\sigma(p)=\pm1$
 define the
 magnetic moments as
 ${\bf S}(p)=\hat{\bf u}(p)\sigma(p)$.
 Then,
 the simplest model is given by a Hamiltonian with nearest-neighbor (NN)
 coupling $J$:
 \begin{equation}
  H_{\rm SI}
   =-3J\sum_{\langle p,p'\rangle}{\bf S}(p)\cdot{\bf S}(p')
   = J \sum_{\langle p,p'\rangle}\sigma(p)\sigma(p').
   \label{Hamil}
 \end{equation}
 We used the relation
 $\hat{\bf u}(p)\cdot\hat{\bf u}(p')=-\frac13$
 for NN spins.
 Because of this anisotropy,
 an effective ferromagnetic model
 \cite{Hertog00}
 is reduced to a fully-frustrated Ising model
 \cite{Villain77,Anderson56}
 for which the ground state consists of tetrahedra satisfying the ``two
 in-two out'' condition (the ice rule \cite{Bernal33}),
 and exhibits Pauling's entropy.
 For later convenience,
 we also define the diamond lattice $\Lambda_{\rm d}$ whose site ($d$)
 sits at the center of each tetrahedron, and introduce a quantity
 $\rho(d):=\eta(d)\sum_{\{p_d\}}\sigma(p_d)/2$.
 $\{p_d\}$ denotes a set of nearest-neighboring sites to $d$,
 and the factor ${\eta(d)}=\pm1$ is assigned to the sublattices of
 $\Lambda_{\rm d}$ to which $d$ belongs.
 $\rho(d)$ is nonzero for tetrahedra not satisfying the ice rule,
 and can be regarded as the charge of point defects
 (i.e., the magnetic monopolelike excitations).
 
 {\it Algorithm:}
 We start with a definition of the local Boltzmann weights, and then
 introduce graphs.
 Let $S^\mu_u$ stand for the 16 states per tetrahedron, where
 $\mu\in[0,4]$ is the number of inward spins, thereby specifying the
 class, and $u$ is the member index in that class
 (see Fig.\ \ref{PYROSG}).
 Then,
 the local Boltzmann weight
 $\omega(S^\mu_u)$
 is independent of $u$,
 and further
 $\omega(S^4_u)=\omega(S^0_u)=z^4$,
 $\omega(S^3_u)=\omega(S^1_u)=z$, and
 $\omega(S^2_u)=1$,
 where
 $z:=\exp(-2/T)$
 (we choose the lowest energy per tetrahedron, $-2J$, as an origin, and
 measure the temperature $T$ in units of $J/\kB$).
 Following the Fortuin-Kasteleyn argument
 \cite{FK72},
 we re-express the system as a superposition of random clusters with
 positive coefficients.
 Because the spin configurations are two in-two out in the ice manifold,
 the transition graphs between any two of them are loop-gas-like.
 We thus expect emerging clusters to form loops at lower $T$
 \cite{Evertz93}.
 This condition is satisfied if only inward-outward pairs of spins {\it can}
 be linked by bonds, just like two Ising spins being energetically
 satisfied in the Swendsen-Wang (SW) algorithm
 \cite{Swendsen87}.
 To give this concept shape,
 we introduce the 10 graphs $G^\nu_v$ (see Fig.\ \ref{PYROSG})
 classified by a number of bonds $\nu\in[0,2]$ ($v$ indexes members in
 the class),
 and consider the following weight decomposition equation
 \cite{Kandel91,Evertz2003}:
 \begin{equation}
  \omega(S^\mu_u)
   =\sum_{\nu,v}\Delta(S^\mu_u,G^\nu_v)W(G^\nu_v),
   \label{WDE}
 \end{equation}
 where
 $W(G^\nu_v)\ge0$ are graph weights to be determined.
 Here,
 we have introduced the factor
 $\Delta(S^\mu_u,G^\nu_v)=0\text{~or~}1$
 to express {\it compatibility} between states and graphs under the
 abovementioned condition:
 For instance, for a given state $S^1_0$, 
 $\Delta(S^1_0,G^1_0)=\Delta(S^1_0,G^1_2)=\Delta(S^1_0,G^1_5)=
 \Delta(S^1_0,G^0_0)=1$,
 and 0 otherwise.
 This factor ensures that the system is decomposed into a mixture of
 independent loops and strings composed of spins directed along their
 lines.
 Once the graph weights satisfying Eq.\ (\ref{WDE}) are given, we can
 then express the conditional probability to assign a graph to the
 tetrahedron in a given state as
 $\Pr(G^\nu_v|S^\mu_u)=\Delta(S^\mu_u,G^\nu_v)W(G^\nu_v)/w(S^\mu_u)$.
 Because Eq.\ (\ref{WDE}) is underdetermined, it is necessary to
 search for a solution guided by the symmetries:
 As the Boltzmann weights are independent of $u$, i.e.,
 $\omega(S^\mu_u)=:\omega^\mu$,
 we also assume that the graph weights are independent of $v$, i.e.,
 $W^\nu:=W(G^\nu_v)$. 
 Then, 
 Eq.\ (\ref{WDE}) is reduced to
 $\omega^\mu=\sum_\nu\Gamma^{\mu\nu}W^\nu$,
 where
 $\Gamma^{\mu\nu}:=\sum_v\Delta(S^\mu_u,G^\nu_v)$
 is independent of $u$.
 The reduced equations are now straightforward to solve;
 we obtain the following solution:
 \begin{equation}
  W^0=z^4,~~~
  W^1=\frac{z-z^4}{3},~~~
  W^2=\frac{3-4z+z^4}{6}. 
   \label{W012}
 \end{equation}
 For the reader's convenience,
 we have summarized the details of our formulation in Supplemental
 Material.

 Here, some comments are in order:
 First,
 in general, the equations such as Eq.\ (\ref{WDE}) do not always
 possess positive solutions for all parameter region,
 and sometimes we should divide the region into parts, in each of which
 a proper choice of graphs is necessary to obtain positive solution.
 Meanwhile, Eq.\ (\ref{W012}) is positive definite, and can be used in
 simulations for all $T$ to generate spin configurations according to
 the Boltzmann weight.
 Second,
 the ergodicity condition obviously holds at finite $T$ because the
 zero-bond graph weight $W^0$ is nonzero for all $S^\mu_u$
 (see Supplemental Material),
 and thus an arbitrary spin configuration can be
 realized with nonvanishing probability.
 At $T=0$,
 any two configurations, as abovementioned, are connected by the
 transition graphs consisting of loops
 \cite{Evertz93}.
 In our cluster decomposition,
 only two-bond graphs $G^2_v$ engage ($W^\nu=\frac12\delta_{\nu2}$),
 so our algorithm reduces to the standard loop algorithm for ice models
 \cite{Barkema98,Yanagawa79}.
 Therefore,
 the updates via loops so created can realize any spin configurations
 also in the ice manifold.
 Third, 
 a short explanation of the improved estimators 
 \cite{Wolff88}
 is given by taking an Ising spin correlation
 ${\cal S}(p,p'):=\langle\sigma(p)\sigma(p')\rangle$
 as an example.
 Like the SW algorithm,
 we decompose the whole system into a set of clusters
 $\{{\cal L}_1,\cdots,{\cal L}_m\}$.
 So,
 we can improve the estimators by taking an average over the $2^m$
 spin configurations by hand which are virtually created by applying
 transition graphs made of $\{\cal L\}$ to a spin configuration in the
 Markov chain.
 Because of the randomness of clusters and the nature of the observable
 $\sigma(p)\sigma(p')$,
 the resulting expression is simply given as
 \begin{equation}
  [\sigma(p)\sigma(p')]_{\rm impr}
   =\delta(p,p'|\{{\cal L}\})\sigma(p)\sigma(p'),
   \label{Simpr}
 \end{equation}
 where
 $\delta(p,p'|\{{\cal L}\})=1$ if two
 sites
 are on the same cluster,
 and 0 otherwise.
 Further,
 $\sigma(p)\sigma(p')$
 only depends on the even-odd parity of the distance between two
 spins
 along the line.
 Because the spin correlation function is the most fundamental quantity,
 we benefit from Eq.\ (\ref{Simpr}) in calculations of other physical
 quantities.
 Indeed,
 we discuss the defect charge correlation
 ${\cal C}(d,d'):=-\langle\rho(d)\rho(d')\rangle$, 
 whose observable is also improved as
 $[\rho(d)\rho(d')]_{\rm impr}
   =\eta(d)\eta(d')\sum_{\{p_d\}\{p_{d'}\}}
   [\sigma(p_d)\sigma(p_{d'})]_{\rm impr}/4$.
 This expression is much more efficient than the enumeration of
 ${\cal C}$ via $\rho(d)$.

 {\it Performance:}
 We shall now assess the performance of our algorithm.
 More explicitly, 
 we estimate equilibrium relaxation times observed in the Metropolis and
 the cluster algorithms, and compare them for various $T$.
 In doing this,
 it is important to choose a proper quantity to characterize the
 relaxation. 
 Our model Eq.\ (\ref{Hamil}) is known not to order at any $T$,
 but exhibits cooperative paramagnetic behavior of diverging susceptibility
 \cite{Ryzhkin05,Jaubert13}.
 Therefore,
 we measure the relaxation times $\tau$ via fluctuations of the uniform
 magnetization
 $M:=\sum_{p}{\bf e}_z\cdot{\bf S}(p)$
 \cite{Jaubert11}.
 First,
 we calculate the time-displaced auto-correlation function of $M$: 
 Define one MC step (MCS) as $|\Lambda_{\rm p}|$
 successive MC trials giving a unit of time for the MC dynamics
 \cite{COMM},
 and denote a measurement of
 $M$ at a $l$th MCS as $M(l)$. 
 The correlation function is then given as
 $A_M(k)=\langle M(k+l)M(l)\rangle$. 
 In Fig.\ \ref{SDT}(a), 
 we provide
 $\bar A_M(k)=A_M(k)/A_M(0)$
 obtained by the two algorithms. 
 The system with a linear dimension $L=64$ ($|\Lambda_{\rm p}|=4L^3$) is
 simulated at $T=0.3$ (circles), 0.5 (diamonds), and 1.0 (triangles). 
 Then,
 we observe a simple relaxation in the Metropolis dynamics,
 so we can evaluate $\tau$ via fitting the data points
 (see blue marks with solid curves). 
 In contrast,
 the relaxation in the cluster dynamics given by red marks is very quick
 so that we cannot evaluate $\tau$ from these plots. 
 To put forward our performance analysis,
 we next employ the statistical-dependence-time (SDT) approach
 \cite{Kikuchi93,Muller73Allen87}
 which only requires statistical quantities, instead of the
 time-displaced correlation and data fitting.
 The SDT is defined as a ratio of two-types of variances $\sigma^2$ and
 $(\delta M)^2_n$: 
 The former is an uniform magnetic susceptibility
 (multiplied by $T$),
 and is algorithm independent.
 However,
 the latter is a variance of statistically independent short-time
 averages with a given bin length $n$: 
 $\langle M_\alpha\rangle_n=\frac{1}{n}\sum_{k=1}^n M_\alpha(k)$
 and 
 $(\delta M)_n^2=\frac{1}{N}\sum_{\alpha=1}^N\langle M_\alpha\rangle_n^2$.
 $N$ is a number of independent runs, and is supposed to be large
 (e.g., $N=2304$ in our calculation).
 Then,
 the ratio,
 $\tau_{\rm dep}:=n(\delta M)_n^2/2\sigma^2$,
 defines SDT, and estimates $\tau$ in the limit $n\nearrow\infty$
 \cite{Kikuchi93}.
 We plot the $n$ dependence of $\tau_{\rm dep}$ in Fig.\ \ref{SDT}(b).
 We find a convergence of the data as bin length increases,
 and further,
 for the Metropolis algorithm,
 the limiting values agree well with the relaxation times estimated from
 $A_M(k)$ (blue dotted lines)
 demonstrating a reliability of the SDT approach
 \cite{Kikuchi93}.
 Meanwhile,
 the cluster algorithm exhibits $\tau_{\rm dep}\simeq\frac12$ (red
 dotted line) that is independent of $n$ and $T$. 
 This implies, assuming single mode relaxation, a vanishing of $\tau$
 for all $T$
 \cite{Kikuchi93}.
 In Fig.\ \ref{SDT}(c), 
 we also check the system-size dependence of $\tau_{\rm dep}$ up to
 $L=64$ at maximum bin lengths $n_{\rm max}(T)$, and confirm its absence
 in these data. 
 Finally, in Fig.\ \ref{SDT}(d), 
 we compare $T$ dependence of $\tau$ estimated via $\tau_{\rm dep}$.
 Although the relaxation time increases as
 $\tau_{\rm M}\propto \exp(2.1/T)$
 for the Metropolis algorithm, it is the constant $\frac12$ indicating
 $\tau_{\rm C}=0$ in the cluster algorithm.
 An energy scale in $\tau_{\rm M}$, $\Delta E\simeq 2.1$, is close to
 the defect creation energy, so that the relaxation depends on point
 defects.
 In contrast, spins bundled in loops and strings can be flipped
 randomly, and thus the cluster algorithm can update them to completely
 new configurations. 
 Consequently, 
 spin-freezing inherent in the single-spin-flip Metropolis algorithm is
 absent in our cluster MC simulations.

 \zuSDT

 {\it Results:}
 We shall now summarize our numerical calculation data; in particular,
 as a hallmark of the ice system, we first demonstrate the emergence of
 pinch points in the spin structure factor
 ${\cal S}_{\perp}({\bf Q})$
 \cite{Youngblood81,Yoshida02,Henley05}.
 This quantity is defined by an average of the squared norm of the
 Fourier component of spins perpendicular to the wavevector ${\bf Q}$,
 i.e., $\langle\|\hat{\bf Q}\times{\bf S}({\bf Q})\|^2\rangle$,
 and gives the diffuse neutron scattering intensity
 (aside from an atomic form factor).
 Although the pinch-point structure observed in experiments
 \cite{Bramwell01PRL}
 has been reproduced in MC calculations
 \cite{Fennell09,Jaubert13},
 system sizes and temperatures simulated are limited to relatively small
 and high values, respectively. 
 Therefore,
 we calculate ${\cal S}_{\perp}({\bf Q})$ for the $L=256$ system at
 temperatures down to $T=0.3$
 (for instance, for {\Dy}, {\LpMax} Dy$^{3+}$ at about {\LowestT} K
 \cite{Hertog00}),
 and then create a high-resolution scattering-intensity color map.
 By comparing it with data for a 4,000-spin system given in the
 loop-update MC study
 \cite{Jaubert13},
 one finds a considerable improvement, and thus a significant development
 in our MC simulation of spin ice.
 Also, 
 we calculate the charge structure factor
 ${\cal C}({\bf Q})=\langle|\rho({\bf Q})|^2\rangle$,
 and provide information on spatial correlations among defects.

 \zuPP

 Following a previous study
 \cite{Bramwell01PRL},
 we scan the plane in the dual space to the periodic lattice
 parametrized by
 ${\bf Q}=(2\pi/a)(h{\bf e}_x+h{\bf e}_y+l{\bf e}_z)$,
 and conventionally denoted as $(h,h,l)$.
 Then,
 reflecting the structure of $\Lambda_{\rm p}$,
 ${\cal S}_{\perp}({\bf Q})$ taken near some special wavevectors plays a
 role in measuring the extent to which the ice rule holds.
 In the left panels of Fig.\ \ref{PP}, 
 we display color maps, for which from bottom to top $T=0.3$, 1.0 and
 2.0.
 At each $T$,
 the data are linearly mapped to colors in the bar with its top and
 bottom colors assigned to the maximum and minimum values.
 At $T=0.3$, we can observe the pinch-point structure at
 (0,0,2), (1,1,1), (2,2,2), and so on,
 but
 it becomes increasing unclear as $T$ increases; this reflects the
 breaking of the ice rule and the increasing of defects.
 This tendency can be recognized also in the charge structure factor: 
 In the right panels of Fig.\ \ref{PP},  
 we give ${\cal C}({\bf Q})$ for the same plane at the same
 temperatures.
 We observe sharp spots at lower $T$ and their diffusion with increasing
 $T$. 
 In particular,
 the one around $(0,0,0)$ reflects long-distance properties indicating
 isotropic short-range charge correlations in real space. 
 Moreover,
 the radius of the spot, a measure of the inverse charge correlation
 length $1/\lambda_{\rm c}$, shortens with the decrease of $T$, and thus
 of the defect number density
 $n_{\rm m}:=\langle\sum_d|\rho(d)|\rangle/|\Lambda_{\rm d}|$.

 To understand the defect distributions,
 the dumbbell model, i.e., the gas of defects interacting via a Coulomb
 potential is useful.
 Castelnovo, Moessner, and Sondhi introduced it into the dipolar spin ice
 \cite{Castelnovo08},
 and, based on the Debye-H\"uckel approximation
 \cite{Debye23},
 argued a screening effect for which they obtained a Debye screening
 length
 $\propto\sqrt{T/\mu_0 n_{\rm m}}$
 ($\mu_0$ is the magnetic permeability)
 \cite{Castelnovo11}.
 We also consider a gas of defects interacting via a Coulomb potential,
 but should take its origin being entropic for Eq.\ (\ref{Hamil}) into
 account, i.e.,
 $V_{\rm ent}\propto T\rho(d)\rho(d')/\|{\bf x}(d)-{\bf x}(d')\|$
 [${\bf x}(d)$ is a position vector of the $d$th site]
 \cite{CP}.
 With $T$ replacing $\mu_0$ as coupling constant,
 their result should read 
 $\lambda_{\rm c}\propto\sqrt{1/n_{\rm m}}$.
 With the structure factors appearing to follow the prediction,
 we shall look more closely at the screening effect:
 Figures\ \ref{FSING}(a) and \ref{FSING}(b) plot ${\cal S}(p,p')$ and
 ${\cal C}(d,d')$ between two sites in the same sublattice;
 their difference vectors are parallel to a primitive axis, i.e.,
 ${\bf x}-{\bf x}'=r{\bf t}_i$.
 Because of the high accuracy of the data,
 $\lambda_{\rm s}$ (a spin correlation length) and $\lambda_{\rm c}$
 can be obtained reliably from these plots, from which the results,
 together with $n_{\rm m}$, are summarized in
 Fig.\ \ref{FSING}(c):
 We then find that
 $\lambda_{\rm s}\simeq\lambda_{\rm c}\propto z^{-1/2}$
 \cite{Sen13}
 and
 $n_{\rm m}\propto z$, which confirms the Debye screening in
 spin ice. 
 Consequently, 
 the Coulomb gas---the emergent picture for spin ice---is efficiently
 simulated via the cluster updates of spins almost satisfying the ice
 rule.
 Finally,
 to shed light on the deconfinement transition from the Coulomb phase to
 the cooperative paramagnetic phase at finite $T$
 \cite{Henley05},
 we perform a finite-size-scaling analysis of $n_{\rm m}(z,L)$ in
 Fig.\ \ref{FSING}(d).
 The scaled quantity $L^{1/\nu}n_{\rm m}(z,L)$ versus the scaled
 variable $L/\xi=Lz^\nu$ for different system sizes is plotted with 
 the exponent $\nu=\frac13$, and reveals good scaling characteristics. 
 This exhibits
 $\xi\propto z^{-1/3}\propto\sqrt[3]{1/n_{\rm m}}$,
 {\it the mean distance among defects}, and the singular part of
 the free-energy density relevant to the deconfinement transition is
 simply given by $f_{\rm s}\propto\xi^{-3}\propto z$
 \cite{Henley05,Powell13}.

 \zuFSING

 {\it Conclusion:}
 We proposed a MC algorithm for spin ice, which decomposes the whole
 system into a mixture of independent loops and strings.
 In contrast to the Metropolis algorithm,
 it is free from spin-freezing;
 we constructed high-resolution color maps of the structure factors
 to demonstrate its efficiency.
 Also,
 we proffered evidence to support Debye screening in spin ice, and
 concluded that the mean distance among defects is relevant to the
 deconfinement transition.
 Apparently,
 an extension of our loop-string algorithm applicable to dipolar spin
 ices is very important.
 Recently, we have evaluated its possible combined use with the
 Metropolis algorithm by which the dipolar interactions are treated
 under the fixed loop-string configurations;
 we shall report on this topic elsewhere
 (see also Ref.\ \cite{Shinaoka2010}).

 The author thanks
 H. Takatsu,
 K. Goto,
 H. Kadowaki,
 K. Chie,
 and 
 Y. Okabe
 for stimulating discussions.
 The main computations were performed using the facilities of Cyberscience
 Center in Tohoku University.
 This work was supported by
 JSPS KAKENHI (26400399).
 
 \newcommand{\AxS}[1]{#1,}
 \newcommand{\AxD}[2]{#1 and #2,}
 \newcommand{\AxM}[2]{#1, and #2,}
 \newcommand{\REF }[4]{#1 {\bf #2}, #3 (#4)}
 \newcommand{\JPSJ}[3]{\REF{J. Phys. Soc. Jpn.\           }{#1}{#2}{#3}}
 \newcommand{\PRL }[3]{\REF{Phys. Rev. Lett.\             }{#1}{#2}{#3}}
 \newcommand{\PRA }[3]{\REF{Phys. Rev.\                  A}{#1}{#2}{#3}}
 \newcommand{\PRB }[3]{\REF{Phys. Rev.\                  B}{#1}{#2}{#3}}
 \newcommand{\PRE }[3]{\REF{Phys. Rev.\                  E}{#1}{#2}{#3}}
 \newcommand{\PRX }[3]{\REF{Phys. Rev.\                  X}{#1}{#2}{#3}}
 \newcommand{\NPB }[3]{\REF{Nucl. Phys.\                 B}{#1}{#2}{#3}}


\begin{thebibliography}{99}
\bibitem{COMM}
	 We denote numbers of lattice points in
	 $\Lambda_{\rm p}$ and 
	 $\Lambda_{\rm d}$
	 as
	 $|\Lambda_{\rm p}|$ and
	 $|\Lambda_{\rm d}|$, 
	 respectively.
\end{thebibliography}
\end{document}